\documentclass[journal]{IEEEtran}
\usepackage{graphicx,color}

\usepackage[OT1]{fontenc}
\usepackage[numbers,sort&compress]{natbib}
\usepackage[cmex10]{amsmath}
\usepackage{amssymb}

\usepackage{bm}
\usepackage{graphicx}
\usepackage{subfigure}
\usepackage{tabularx}
\usepackage{arydshln}
\usepackage{mathtools}
\usepackage{multirow}
\usepackage{algorithm,algorithmic}
\usepackage{url}
\usepackage{listings}
\usepackage{comment}
\usepackage{braket}
\usepackage{color}
\usepackage{hyperref}
\usepackage[absolute,overlay]{textpos}

\def\NCH{N_{\mathrm{CH}}}
\def\NAP{N_{\mathrm{AP}}}
\def\NB{N_{\mathrm{B}}}
\def\etal{et~al.~}

\makeatletter
\def\ps@IEEEtitlepagestyle{%
  \def\@oddfoot{\mycopyrightnotice}%
  \def\@oddhead{\hbox{}\@IEEEheaderstyle\leftmark\hfil\thepage}\relax
  \def\@evenhead{\@IEEEheaderstyle\thepage\hfil\leftmark\hbox{}}\relax
  \def\@evenfoot{}%
}
\def\mycopyrightnotice{%
  \begin{minipage}{\textwidth}
  \centering \scriptsize
\textcopyright 2023 IEEE. Personal use of this material is permitted.
  Permission from IEEE must be obtained for all other uses, in any current or future
  media, including reprinting/republishing this material for advertising or promotional
  purposes, creating new collective works, for resale or redistribution to servers or
  lists, or reuse of any copyrighted component of this work in other works.
  DOI: \href{https://doi.org/10.1109/TQE.2023.3293452}{10.1109/TQE.2023.3293452}
  \end{minipage}
}
\makeatother
\begin{document}
\title{Qubit Reduction and Quantum Speedup for Wireless Channel Assignment Problem}

\author{Yuki~Sano,~Masaya~Norimoto,~\IEEEmembership{Graduate Student Member,~IEEE}, and Naoki~Ishikawa,~\IEEEmembership{Senior~Member,~IEEE}.\thanks{Y.~Sano, M.~Norimoto, and N.~Ishikawa are with the Faculty of Engineering, Yokohama National University, 240-8501 Kanagawa, Japan (e-mail: ishikawa-naoki-fr@ynu.ac.jp). This research was partially supported by the Japan Society for the Promotion of Science (JSPS) KAKENHI (Grant Number 22H01484) and the MITOU Program, Information-technology Promotion Agency, Japan.}}

\maketitle
\TPshowboxesfalse
\begin{textblock*}{\textwidth}(45pt,10pt)
\footnotesize
\centering
Accepted for publication in IEEE Transactions on Quantum Engineering. This is the author's version which has not been fully edited and content may change prior to final publication. Citation information: DOI 10.1109/TQE.2023.3293452
\end{textblock*}

\begin{abstract}
In this paper, we propose a novel method of formulating an NP-hard wireless channel assignment problem as a higher-order unconstrained binary optimization (HUBO), where the Grover adaptive search (GAS) is used to provide a quadratic speedup for solving the problem. The conventional method relies on a one-hot encoding of the channel indices, resulting in a quadratic formulation. By contrast, we conceive ascending and descending binary encodings of the channel indices, construct a specific quantum circuit, and derive the exact numbers of qubits and gates required by GAS. Our analysis clarifies that the proposed HUBO formulation significantly reduces the number of qubits and the query complexity compared with the conventional quadratic formulation. This advantage is achieved at the cost of an increased number of quantum gates, which we demonstrate can be reduced by our proposed descending binary encoding.
\end{abstract}


\IEEEpeerreviewmaketitle

\section{Introduction}

\IEEEPARstart{T}{HE} radio spectrum is a limited resource, whereas global mobile traffic is expected to increase 77-fold between 2020 and 2030 \cite{union2015imt}.
It is vital to support as many user terminals (UTs) as possible within the limited spectrum.
To achieve efficient utilization of the spectrum, the problem of determining which channel access points (APs) should use is called the \textit{channel assignment problem (CAP)}.
It is known that CAP is equivalent to the generalized graph coloring problem \cite{hale1980frequency}, which is classified as NP-hard.
CAP incurs a combinatorial explosion as the numbers of APs and UTs increase.
Numerous greedy algorithms have been proposed \cite{ramanathan1999unified} to obtain practical solutions within a limited amount of time, but none of them can guarantee that an optimal solution will be obtained.

In solving the NP-hard CAP, quantum annealing (QA) and the coherent Ising machine (CIM) have been demonstrated to be effective solutions.
QA was used to solve the graph coloring problem \cite{kwok2020graph}.
It has been applied to the dynamic spectrum allocation problem and found to obtain a practical solution in a short time \cite{wang2016quantum,saito2021evaluating}.
CIM is another type of physical annealing system that outperforms QA for some specific problems \cite{hamerly2019experimental}.
Its application to the graph coloring problem is found in \cite{inaba2022potts}.
Hasegawa \etal formulated CAP as a quadratic unconstrained binary optimization (QUBO) problem and demonstrated that CIM was capable of maximizing the channel capacity \cite{ito2020highspeed,hasegawa2021optimization}.
It also succeeded in maximizing the total throughput of UTs in a dense Wi-Fi network \cite{kurasawa2021highspeed}.
In those conventional studies, both QA and CIM were demonstrated to be effective technologies for industrial applications.
One remaining concern is that neither can guarantee their optimality, which is the same as other metaheuristics, including the greedy approach.

One promising approach that can guarantee global optimality is a quantum algorithm \cite{botsinis2019quantum}, \textit{Grover adaptive search (GAS)} \cite{gilliam2021grover,gilliam2020optimizing}, which relies on quantum superposition, entanglement, and amplitude amplification.
Unlike QA and CIM, GAS supports higher-order unconstrained binary optimization (HUBO) problems in addition to QUBO problems.
As with other Grover-type quantum algorithms \cite{grover1996fast,boyer1998tight,durr1999quantum,tezuka2022grover}, GAS provides quadratic speedups for solving QUBO and HUBO problems.
That is, it achieves the query complexity of $O(\sqrt{2^n})$ with $n$ binary variables, while the classic exhaustive search requires $O(2^n)$ function evaluations.
Following the innovative GAS proposed by Gilliam \etal \cite{gilliam2021grover}, an extension method that supports HUBO with real-valued coefficients was proposed in \cite{norimoto2023quantum}.
In addition, heuristic efforts and problem-specific attributes can be used to further reduce the constant overhead of GAS \cite{giuffrida2022engineering,yukiyoshi2022quantum}.
The major challenge with quantum algorithms is the correction of quantum errors induced by noise \cite{fujii2016noise}.
Quantum algorithms, including GAS, require many physical qubits for quantum error correction \cite{gidney2021how} and share a common issue in terms of feasibility.

To reduce the number of required qubits and improve the feasibility, sophisticated formulations have been considered in the conventional studies: one-hot encoding and binary encoding.
One-hot encoding indicates an index by a one-hot vector, such as $2 \rightarrow [0~1~0~0]$ and $4 \rightarrow [0~0~0~1]$, whereas binary encoding indicates an index by its binary representation, such as $2 \rightarrow [0~1]$ and $4 \rightarrow [1~1]$.
In quantum computation, these encoding methods were first considered by Sawaya \etal \cite{sawaya2020resourceefficient} who aimed at improving the efficiency of quantum simulations.\footnote{The encoding method of Sawaya \etal was published on September 29, 2019 \cite{sawaya2020resourceefficient}, which would be considered the first innovation in quantum computation. Note that in the context of the traveling salesman problem, Lidd's binary encoding of city indices was proposed in 1991 \cite{larranaga1999genetic}, but this contribution was limited to cross-over and mutation operations for genetic algorithms.}
Then, Fuchs \etal used the one-hot and binary encodings for the weighted Max-Cut problem and showed that the number of required qubits could be reduced by one logarithmic order \cite{fuchs2021efficient}.
Glos \etal used these encodings and formulated the traveling salesman problem as a HUBO problem \cite{glos2022spaceefficient}.
Similarly, Tabi \etal used the binary encoding for the graph coloring problem and showed a reduction in the number of qubits \cite{tabi2020quantum}.
Note that all the above conventional studies \cite{fuchs2021efficient,glos2022spaceefficient,tabi2020quantum} considered
the quantum approximate optimization algorithm (QAOA) for noisy intermediate-scale quantum computers rather than the GAS for fault-tolerant quantum computers.

Classically, the main approach to solving a HUBO problem is to transform it into a QUBO problem, and then use the semidefinite relaxation technique \cite{luo2010semidefinite} to obtain a good but suboptimal solution in polynomial time.
This transformation is termed \textit{quadratization} \cite{rosenberg1975reduction,anthony2017quadratic}, and it involves an increase in the number of binary variables, resulting in an exponential increase in the search space size.
By contrast, both QAOA and GAS can deal directly with HUBO problems owing to the interaction of multiple qubits, and do not impose the quadratic constraint required for the semidefinite relaxation.

Against this background, we propose a novel method of formulating CAP as a HUBO problem and analyze the numbers of qubits and gates required by GAS; this is a first attempt in the literature.
The major contributions of this paper are summarized as follows.
\begin{enumerate}
    \item We represent the channel indices as binary numbers and formulate the NP-hard CAP as a HUBO problem, whereas all the conventional studies \cite{ito2020highspeed, kurasawa2021highspeed, hasegawa2021optimization} formulated CAP as a QUBO problem. Compared with the conventional QUBO formulation, the number of qubits required by GAS is significantly reduced by one logarithmic order, where a specific quantum circuit is constructed without any black-box quantum oracle.
    \item For the conventional QUBO and the proposed HUBO formulations, we derive the exact numbers of required quantum gates and clarify corresponding asymptotic orders. Unlike the conventional studies \cite{fuchs2021efficient,glos2022spaceefficient,tabi2020quantum}, we conceive novel ascending and descending assignments for the binary encoding. The latter significantly reduces the number of quantum gates.
    \item We confirm that the theoretical speedup by GAS is also effective for CAP. Here, the proposed HUBO formulation significantly reduces the query complexity in the classical domain compared with the conventional QUBO owing to the reduced search space caused by the reduced number of qubits.
\end{enumerate}

The remainder of this paper is organized as follows.
In Section~\ref{sec:gas}, we introduce the conventional GAS that supports real-valued coefficients.
Section~\ref{sec:conv} is a review of the conventional QUBO formulation for CAP, while in Section~\ref{sec:prop}, we propose our HUBO formulation.
For both formulations, algebraic and numerical evaluations are given in Section~\ref{sec:comp}.
Finally, in Section~\ref{sec:conc}, we conclude this paper.
\begin{table}[tbp]
	\centering
	\caption{List of important mathematical symbols\label{table:sym}}
	\begin{tabular}{lll}
	    $\mathbb{B}$ & & Binary numbers \\
	    $\mathbb{R}$ & & Real numbers \\
	    $\mathbb{C}$ & & Complex numbers \\
	    $\mathbb{Z}$ & & Integers \\
	    $j$ & $\in \mathbb{C}$ & Imaginary number \\
	    $\NAP$ & $\in \mathbb{Z}$ & Number of APs \\
		$\NCH$ & $\in \mathbb{Z}$ & Number of channels \\
		$\NB$ & $\in \mathbb{Z}$ & Number of bits required to represent $\NCH$ \\
		 & &channel indices, i.e., $\NB = \lceil \log_2 \NCH \rceil$\\
		$B_{\mathrm{w}}$ & $\in \mathbb {R}$ & Bandwidth \\
		$P$ & $\in \mathbb {R}$ & Transmit power \\
		$x$ & $\in \mathbb {B}$ & Binary variable \\
		$E(x)$ & $\in \mathbb{Z}$ & Conv. objective function for QUBO \\
		$E'(x)$ & $\in \mathbb{Z}$ & Prop. objective function for HUBO \\
		$G$ & $\in \mathbb{Z}$ & Number of gates required by conv. QUBO \\
		$G'$ & $\in \mathbb{Z}$ & Number of gates required by prop. HUBO \\
		$i, k$ & $\in \mathbb {Z}$ & Index of APs or GAS iterations \\
		$c, l$ & $\in \mathbb {Z}$ & Index of channels \\
		$d_{iu}$ & $\in \mathbb {R}$ & Physical distance between $i$th AP and $u$th UT \\
		$\alpha$ & $\in \mathbb {R}$ & Attenuation coefficient \\
		$n$ & $\in \mathbb{Z}$ & Number of QUBO variables in $E(x)$ \\
		$m$ & $\in \mathbb{Z}$ & Number of qubits required to encode $E(x)$ \\
		$n'$ & $\in \mathbb{Z}$ & Number of HUBO variables in $E'(x)$ \\
		$m'$ & $\in \mathbb{Z}$ & Number of qubits required to encode $E'(x)$ \\
		$y_i$ & $\in \mathbb{Z}$ & Threshold that is adaptively updated by GAS \\
		$L_i$ & $\in \mathbb{Z}$ & Number of Grover operators \\
	\end{tabular}
\end{table}
Italicized symbols represent scalar values, and bold symbols represent vectors and matrices.
Table~\ref{table:sym} summarizes the important mathematical symbols used in this paper.

\section{Grover Adaptive Search \cite{gilliam2021grover,norimoto2023quantum}\label{sec:gas}}
In this section, we introduce the innovative GAS algorithm proposed by Gilliam \etal \cite{gilliam2021grover} and its modification to support real-valued coefficients \cite{norimoto2023quantum}.
GAS is a quantum algorithm that solves QUBO and HUBO problems owing to its efficient construction method for the corresponding quantum circuit.
It requires $n+m$ qubits \cite{gilliam2021grover}\footnote{In practice, some ancillae are required, which is detailed in Section~\ref{sec:comp}.}, where $n$ is the number of binary variables and $m$ is the number of qubits to encode the objective function value using the two's complement representation. Therefore, $m$ is the smallest integer that satisfies the following two inequalities:
\begin{align}
-2^{m-1} \leq a < 2^{m-1} \label{eq:m1}
\end{align}
and
\begin{align}
-2^{m-1} \leq \min[E(x)] \leq \max[E(x)] < 2^{m-1}. \label{eq:m2}
\end{align}
Here, $a$ is an arbitrary coefficient included in the objective function $E(x)$, which is a quadratic or higher-order polynomial function of $n$ binary variables $(x_1,\cdots,x_n)$.
GAS minimizes the objective function $E(x)$ by two steps: (1) create an equal superposition of $2^n$ quantum states and (2) amplify the states of interest where the values of the objective function become smaller than the current minimum.
If the objective function $E(x)$ contains real-valued coefficients, multiple states may be amplified depending on the value, resulting in a superposition of approximated integers \cite{gilliam2021grover}.
That is, there is a slight probability that inappropriate states will be observed, in which case, GAS may not work.
This inappropriate behavior can be avoided if the algorithm calibrates the objective function value to the correct one in the classical domain.
At the sacrifice of additional query complexity in the classical domain, the modified GAS \cite{norimoto2023quantum} supports real-valued coefficients of QUBO and HUBO functions.

\begin{figure}[tbp]
	\centering
    \includegraphics[clip, scale=0.58]{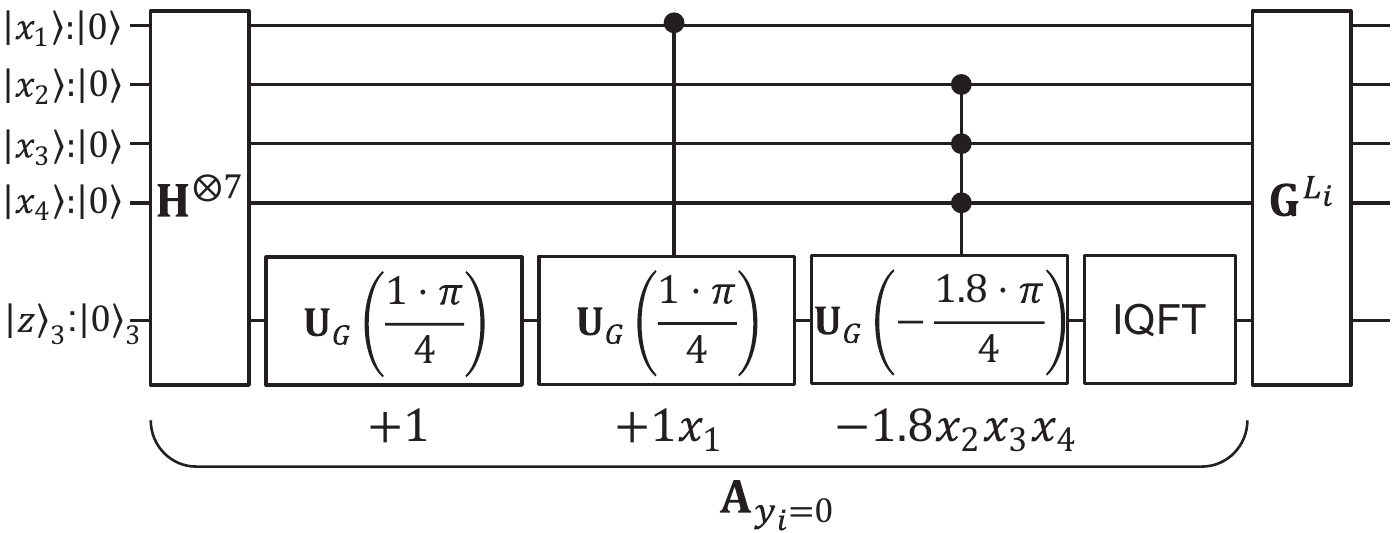}
	\caption{Quantum circuit corresponding to $E(x)=1+x_1-1.8x_2x_3x_4$ and $y_i= 0$. \label{fig:circuit-real-co}}
\end{figure}
In \cite{gilliam2021grover}, Gilliam \etal proposed an efficient method to construct a quantum circuit corresponding to a binary polynomial function $E(x)$.
We introduce this construction method using the example quantum circuit shown in Fig.~\ref{fig:circuit-real-co}, corresponding to $E(x)=1+x_1-1.8x_2x_3x_4$.
Specifically, at an iteration index $i$, a state preparation operator $\mathbf{A}_{y_i}$ calculates values $E(x) - y_i$ corresponding to $2^n$ states, where $y_i$ is the current minimum.
Here, the two's complement is used to represent negative values.
The operator $\mathbf{A}_{y_i}$ is composed of the inverse quantum Fourier transform (IQFT),
the Hadamard gates
\begin{align}
    \mathbf{H} = \frac{1}{\sqrt{2}} \begin{bmatrix}
        +1 & +1 \\
        +1 & -1
    \end{bmatrix},
\end{align}
\begin{align}
    \mathbf{H}^{\otimes n} =
    \underbrace{\mathbf{H} \otimes \cdots \otimes \mathbf{H}}_{n~\textrm{gates}}
\end{align}
that creates an equal superposition, the phase gate
\begin{align}
    \mathbf{R}(\theta) = \begin{bmatrix}
        1 & 0 \\ 0 & e^{j\theta}
    \end{bmatrix},
\end{align}
and the unitary operator \cite{gilliam2021grover}
\begin{align}
    \mathbf{U}_{\mathrm{G}}(\theta) = \mathbf{R}(2^{m-1} \theta)
    \otimes \mathbf{R}(2^{m-2} \theta) \otimes \cdots \otimes \mathbf{R}(2^{0} \theta),
    \label{Ug-theta}
\end{align}
where we have $\theta=2\pi a/2^m$.
The unitary operator $\mathbf{U}_{\mathrm{G}}(\theta)$ corresponds to a coefficient $a$ in the objective function, and it is controlled by a set of qubits corresponding to binary variables.
For example, as shown in Fig.~\ref{fig:circuit-real-co}, the term $+1$ corresponds to the gate $\mathbf{U}_{\mathrm{G}}(+1 \cdot \pi / 4)$ and the term $-1.8x_2x_3x_4$ corresponds to the gate $\mathbf{U}_{\mathrm{G}}(-1.8 \cdot \pi / 4)$ controlled by $(\ket{x_2}, \ket{x_3}, \ket{x_4})$.
The states of interest, in which the values $E(x) - y_i$ become negative, can be identified by focusing on the beginning of $m$ qubits since we use the two's complement.
Then, the oracle operator $\mathbf{O}$ has a single Z gate
\begin{align}
    \mathbf{Z} = \begin{bmatrix}
    1 & 0\\
    0 & -1
    \end{bmatrix}
\end{align}
to invert the phase of the negative states, i.e.,
\begin{align}
    \mathbf{O} = \underbrace{\mathbf{I}_2 \otimes \cdots \otimes \mathbf{I}_2}_{n~\textrm{qubits}} \otimes \mathbf{Z} \otimes
    \underbrace{\mathbf{I}_2 \otimes \cdots \otimes \mathbf{I}_2}_{m-1~\textrm{qubits}}.
\end{align}
The Grover operator $\mathbf{G}=\mathbf{A}_{y_i}\mathbf{D}\mathbf{A}_{y_i}^{\mathrm{H}} \mathbf{O}$ amplifies only such states and is composed of $\mathbf{A}_{y_i}$, $\mathbf{O}$, and the Grover diffusion operator $\mathbf{D}$ \cite{gilliam2021grover}.

In GAS, to amplify the states of interest, the Grover operator $\mathbf{G}$ is applied $L_i$ times, which is a uniform random value.
After that, the quantum states are measured, and a solution $(x_1, \cdots, x_n)$ is obtained.
To calibrate the mismatch induced by the real-valued coefficients, the objective function value $y=E(x)$ is evaluated in the classical domain, and is compared with the current minimum value $y_i$.
If the value is smaller than $y_i$, GAS updates the solution and the current minimum, and sets $k=1$.
Otherwise, GAS sets $k=\min{\{\lambda k,\sqrt{2^n}}\}$ using a scalar $\lambda = 8/7$ \cite{boyer1998tight,baritompa2005grover}.
The termination condition could be based on the sum of the number of Grover operators, the number of classical iterations, or the number of iterations in which the objective value does not improve.
The above algorithm is summarized in Algorithm~\ref{alg:real-gas}.
\begin{algorithm}[tb]
    \caption{GAS supporting real-valued coefficients \cite{gilliam2021grover,norimoto2023quantum}.\label{alg:real-gas}}
    \begin{algorithmic}[1]
        \renewcommand{\algorithmicrequire}{\textbf{Input:}}
        \renewcommand{\algorithmicensure}{\textbf{Output:}}
        \REQUIRE $E(x):\mathbb{B}^n\rightarrow\mathbb{R}, \lambda=8/7$
        \ENSURE $x_1, \cdots, x_n$
        \STATE {Uniformly sample $x_1, \cdots, x_n$ and set $y_0=E(x)$}.
        \STATE {Set $k = 1$ and $i = 0$}.
        \REPEAT
        \STATE\hspace{\algorithmicindent}{Let $L_i$ be a random number from $0$ to $\lceil k-1 \rceil$}.
        \STATE\hspace{\algorithmicindent}{Evaluate $\mathbf{G}^{L_i} \mathbf{A}_{y_i} \Ket{0}_{n+m}$ and obtain $x_1', \cdots, x_n'$}.
        \STATE\hspace{\algorithmicindent}{Evaluate $y=E(x')$ in the classical domain}. \COMMENT{This is an additional step to support real coefficients}
        \hspace{\algorithmicindent}\IF{$y<y_i$}
        \STATE\hspace{\algorithmicindent}{Update the solution $x_1 = x_1'$, $\cdots$, $x_n=x_n'$}.
        \STATE\hspace{\algorithmicindent}{Set $y_{i+1}=y$ and $k=1$}.
        \hspace{\algorithmicindent}\ELSE{\STATE\hspace{\algorithmicindent}{Set $y_{i+1}=y_i$ and $k=\min{\{\lambda k,\sqrt{2^n}}\}$}}.
        \ENDIF
        \STATE{$i=i+1$}.
        \UNTIL{a termination condition is met}.
    \end{algorithmic}
\end{algorithm}

\section{Conventional QUBO Formulation with One-Hot Encoding \cite{ito2020highspeed, kurasawa2021highspeed, hasegawa2021optimization} \label{sec:conv}}
In this section, we review a conventional QUBO formulation for CAP, which relies on the one-hot encoding of channel indices.
\begin{figure}[tbp]
	\centering
    \includegraphics[clip, scale=0.66]{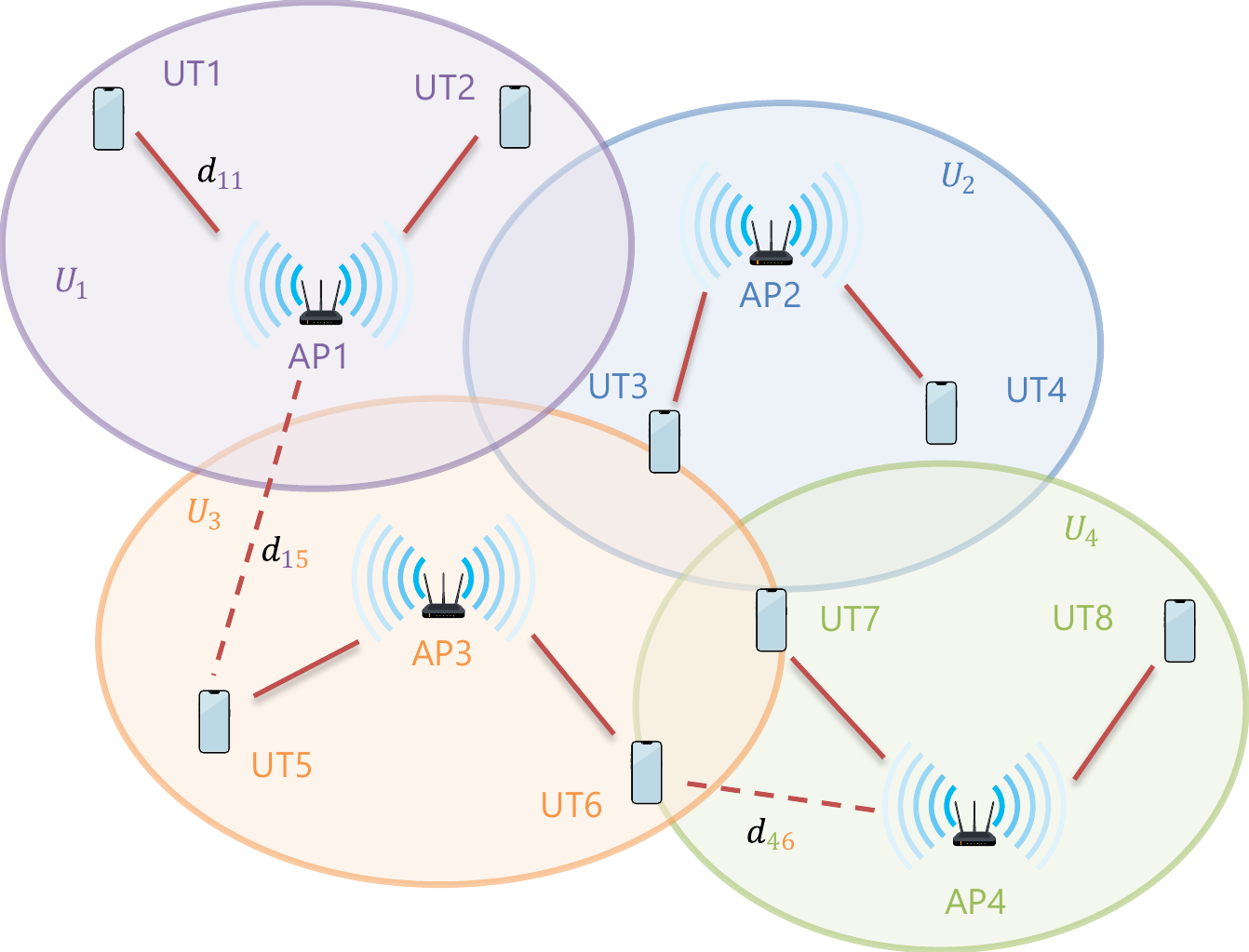}
	\caption{System model of CAP, where the number of APs is $\NAP=4$ and the number of channels is $\NCH=3$.\label{fig:model}}
\end{figure}
Let the number of APs be $\NAP$ and the number of channels be $\NCH$, and we assume that the condition $\NCH < \NAP$ holds.\footnote{When $\NCH \geq \NAP$, the optimal solution for CAP is obvious.}
Fig.~\ref{fig:model} shows an example system setup of CAP, where we consider $\NAP=4$ APs, $\NCH=3$ channels, and eight UTs.
Here, $U_i$ denotes a set of UTs associated with the $i$th AP for $1 \leq i \leq \NAP$. In Fig.~\ref{fig:model}, two UTs are associated with each AP.
The scalar constant $d_{iu}$ denotes the physical distance between the $i$th AP and $u$th UT.
The search space size is calculated as
\begin{align}
    C_{\mathrm{CAP}} =
    {\NCH}^{\NAP},\label{eq:searchspace}
\end{align}
which is not clarified in \cite{ito2020highspeed, kurasawa2021highspeed, hasegawa2021optimization}.

In the conventional studies \cite{ito2020highspeed, hasegawa2021optimization,kurasawa2021highspeed}, the goal is to maximize the total capacity of the network
\begin{align}
\sum_{i=1}^{\NAP}
B_{\mathrm{w}} \log_2\left({1 + \frac{S_i}{I_i + N_i}}\right),\label{eq:C}
\end{align}
where $S_i$ is the power of the received signal of interest at $i$th AP, $I_i$ is the interference power from other APs, $N_i$ is the noise power, and $B_{\mathrm{w}}$ is the bandwidth.
In the following, we consider maximization of the total capacity. Although some studies \cite{ito2020highspeed, hasegawa2021optimization} considered maximizing the sum of signal-to-interference-plus-noise ratios (SINRs), the essential formulation approach below is identical with the conventional studies \cite{ito2020highspeed, hasegawa2021optimization,kurasawa2021highspeed}.
Maximizing \eqref{eq:C} is equivalent to minimizing
\begin{align}
    -\sum_{i=1}^{\NAP} \log_2 \left(1 + \frac{S_i}{I_i}\right),
    \label{eq:IiSi}
\end{align}
which is considered as an objective function.
Then, we prepare $\NAP \NCH$ binary variables defined by \cite{ito2020highspeed,hasegawa2021optimization}
\begin{align}
x_{ic} &=
  \begin{cases}
    1\hphantom & \text{the $i$th AP uses the $c$th channel}\\
    0 & \text{otherwise}
  \end{cases}\label{eq:xij}
\end{align}
for $1 \leq i \leq \NAP$ and $1 \leq c \leq \NCH$.
Since the $i$th AP uses a single channel, $[x_{i1} ~ \cdots ~ x_{i\NCH}] \in \mathbb{B}^{\NCH}$ becomes a one-hot vector; this is also known as one-hot encoding.

Assuming a simplified path-loss channel model, the received signal power of interest at $i$th AP is represented by
\begin{align}
\sum_{u\in U_i} P d_{iu}^{-\alpha}
\label{eq:siu}
\end{align}
where $P$ denotes the transmit power, $U_i$ denotes a set of UTs associated with the $i$th AP, $d_{iu}$ denotes the distance between the $i$th AP and $u$th UT, and $\alpha$ denotes an attenuation coefficient.
Similarly, the total interference at $i$th AP from other APs is represented by
\begin{align}
\sum_{c=1}^{\NCH} \sum_{k=1}^{\NAP}\sum_{l=1}^{\NCH}
\left(
\sum_{v\in U_k} P d_{iv}^{-\alpha}
\right)
\delta_{cl}(1 - \delta_{ik}) x_{ic}x_{kl}, \label{eq:iiu}
\end{align}
where $\delta_{cl}$ denotes the Kronecker delta defined by
\begin{align}
\delta_{cl} =
  \begin{cases}
    1\hphantom & \text{if $c = l$}\\
    0 & \text{otherwise.}
  \end{cases}
\end{align}
That is, the term $\delta_{cl}(1 - \delta_{ik}) x_{ic}x_{kl}$ becomes $1$ if a pair $(i, k)$ of different APs use the same channel $c=l$, and $0$ otherwise.

From \eqref{eq:siu} and \eqref{eq:iiu}, minimizing \eqref{eq:IiSi} is equivalent to minimizing
\begin{align}
E_1(x) =
\sum_{i=1}^{\NAP-1} \sum_{c=1}^{\NCH} \sum_{k=i+1}^{\NAP} \sum_{l=1}^{\NCH}
D_{ik} \delta_{cl}(1 - \delta_{ik}) x_{ic}x_{kl}, \label{eq:E1}
\end{align}
where we have
\begin{align}
D_{ik} = C_{ik} -C_{\mathrm{min}} + \epsilon, \label{eq:Dik}
\end{align}
and
\begin{align}
    C_{ik} = -&\log_2\left( 1 +
\frac{\sum_{u\in U_i} d_{iu}^{-\alpha}}{\sum_{v\in U_k} d_{iv}^{-\alpha}}
\right) \nonumber \\
-&
\log_2\left( 1 +
\frac{\sum_{u\in U_k} d_{ku}^{-\alpha}}{\sum_{v\in U_i} d_{kv}^{-\alpha}}
\right).
\label{eq:Cik}
\end{align}
Here, $\epsilon$ is a small constant such as $\epsilon=0.01$, and $C_{\mathrm{min}}$ is the minimum of $C_{ik}$ for $1 \leq i < k \leq \NAP$.
Both parameters $\epsilon$ and $C_{\mathrm{min}}$ lead to $D_{ik} > 0$.
In addition, each AP uses a single channel.
This constraint can be given by \cite{ito2020highspeed,hasegawa2021optimization,kurasawa2021highspeed}
\begin{align}
E_2(x) = \sum_{i=1}^{\NAP}\left(\sum_{c=1}^{\NCH} x_{ic} - 1\right)^2. \label{eq:E2}
\end{align}
Overall, from \eqref{eq:E1} and \eqref{eq:E2}, CAP is formulated as \cite{ito2020highspeed,hasegawa2021optimization,kurasawa2021highspeed}
\begin{align}
\begin{aligned}
\min_{x} \quad & E(x) = E_1(x) + w E_2(x) \\
\textrm{s.t.} \quad & x_{ic} \in \mathbb{B}
\end{aligned}
\label{eq:E}
\end{align}
for $1 \leq i \leq \NAP$ and $1 \leq c \leq \NCH$.
Here, $w$ is a penalty parameter for satisfying the constraint of $E_2(x)$.
The objective function $E(x)$ is a quadratic function using the one-hot encoding of channel indices.
A specific example of QUBO formulation is given in Appendix~\ref{app:exqubo}.

\section{Proposed HUBO Formulation with Binary Encoding\label{sec:prop}}
As reviewed in Section~\ref{sec:conv}, the conventional QUBO formulation relies on the one-hot encoding.
In this section, a novel HUBO formulation using the binary encoding is proposed, which is solved by the GAS introduced in Section~\ref{sec:gas}.

As with the original QUBO formulation \cite{ito2020highspeed, kurasawa2021highspeed, hasegawa2021optimization}, the goal is to maximize the total capacity \eqref{eq:C} of the network.
The original formulation requires $\NAP \NCH$ binary variables $x_{ic}$ to represent the state where the $i$th AP uses the $c$th channel.
That is, the search space size of the QUBO formulation is calculated as
\begin{align}
    C_{\mathrm{QUBO}} = 2^{\NAP \NCH},
\end{align}
which is larger than the original search space $C_{\mathrm{CAP}} = {\NCH}^{\NAP}$ of \eqref{eq:searchspace}.
According to \eqref{eq:searchspace}, it can be expected that the minimum number of required binary variables is calculated as
\begin{align}
    \log_2 C_{\mathrm{CAP}} = \NAP \log_2{\NCH}.
    \label{eq:log2CCAP}
\end{align}

In the following, we conceive an optimal formulation that can achieve the minimum number \eqref{eq:log2CCAP}, resulting in the minimum size of search space.
Since $\log_2{\NCH}$ is included in \eqref{eq:log2CCAP}, it can be expected that the channel index should be represented by a binary number.
The number of bits $\NB$ required to represent $\NCH$ channel indices is
\begin{align}
\NB = \lceil \log_{2} \NCH \rceil \label{eq:R}
\end{align}
because we have $2^{\NB-1} < \NCH \leq 2^{\NB}$.
Since $\NCH$ is not limited to a power of two, it may allocate up to one bit more than the original search space, which is inefficient.

Here, we propose an encoding method of exploiting the unallocated bit and reducing the number of terms included in a new objective function.
Specifically, we assign $\NB$-bit sequences, starting from all zeros, to the ascending channel indices $c=1, 2, \cdots, \NCH$, which is referred to as \textit{ascending binary encoding}.
In this encoding method, given the channel index $c$,
an $\NB$-bit sequence is defined by
\begin{align}
[b_{c1}~b_{c2}~\cdots~b_{c\NB}] = [c - 1]_2, \label{eq:c-1}
\end{align}
where $[ \cdot ]_2$ denotes the decimal to binary conversion.
By contrast, in another method, we assign the bit sequences to the descending channel indices $c=\NCH, \NCH-1, \cdots, 1$, which is referred to as \textit{descending binary encoding}.
In this descending binary encoding,
an $\NB$-bit sequence is defined by
\begin{align}
[b_{c1}~b_{c2}~\cdots~b_{c\NB}] = [\NCH - c + 1]_2.
\end{align}

For both encoding methods, the state where the $i$th AP uses the $c$th channel is represented by
\begin{align}
\delta'_{ic}(x) = \prod_{r=1}^{\NB} (1 - b_{cr} + (2b_{cr} - 1) x_{ir}), \label{eq:dic}
\end{align}
where we have $\NAP \NB$ binary variables $x_{ir}$ for $1 \leq i \leq \NAP$ and $1 \leq r \leq \NB$.
Note that $\delta'_{ic}(x)$ is a function of order $\NB$.

In the proposed encoding methods, the interference between APs is represented by
\begin{align}
E_1'(x) = \sum_{i=1}^{\NAP-1} \sum_{k=i+1}^{\NAP} D_{ik} \sum_{c=1}^{\NCH}
\delta'_{ic}(x) \delta'_{kc}(x), \label{eq:E1'}
\end{align}
which should be minimized.
Additionally, when $\NCH<2^{\NB}$, we add
\begin{align}
E_2'(x) = \sum_{i=1}^{\NAP} \sum_{c=\NCH+1}^{2^{\NB}} \delta'_{ic}(x) \label{eq:E2'}
\end{align}
due to the constraint that each AP does not use nonexistent channels.
Overall, from \eqref{eq:E1'} and \eqref{eq:E2'}, the proposed HUBO formulation is given by
\begin{align}
\begin{aligned}
\min_{x} \quad & E'(x) = E_1'(x) + w' E_2'(x) \\
\textrm{s.t.} \quad & x_{ir} \in \mathbb{B},
\end{aligned}
\label{eq:E'}
\end{align}
where $w'$ is a penalty parameter for satisfying the constraint of $E_2'(x)$.
The proposed objective function contains terms of the order $2\NB$ at most.
That is, it is a unique HUBO formulation that can be solved by GAS.

\begin{table}[tb]
	\centering
	\caption{Proposed ascending and descending binary encodings.\label{table:ch}}
	\small
	\begin{tabular}{lll}
	    \hline
		Index & Binary encoding (asc.) & Binary encoding (desc.)\\
		$c$ & $[b_{c1} ~ b_{c2}],~~\delta'_{ic}(x)$ & $[b_{c1} ~ b_{c2}],~~\delta'_{ic}(x)$\\
		\hline
		$1$ & $[0 ~ 0],~~(1-x_{i1})(1-x_{i2})$ & $[1 ~ 1],~~x_{i1}x_{i2}$ \\
		$2$ & $[0 ~ 1],~~(1-x_{i1})x_{i2}$ & $[1 ~ 0],~~x_{i1}(1-x_{i2})$ \\
		$3$ & $[1 ~ 0],~~x_{i1}(1-x_{i2})$ & $[0 ~ 1],~~(1-x_{i1})x_{i2}$ \\
		\hline
	\end{tabular}
\end{table}
Let us check a specific example in which $\NAP = 4$, $\NCH = 3$, and $\NB=\lceil \log_2{3} \rceil=2$.
Table~\ref{table:ch} provides the mapping between the channel indices $c = 1, 2, 3$, the binary sequence $[b_{c1} ~ b_{c2}]$, and the function $\delta'_{ic}(x)$, where both ascending and descending binary encodings are considered.
The number of terms in $(1-x_{i1})(1-x_{i2})$ is four, while it is reduced to one in $x_{i1}x_{i2}$.
The corresponding HUBO formulations are given in \eqref{eq:E_prop_a} and \eqref{eq:E_prop_d} of Appendix~\ref{app:exhubo}.
\begin{figure*}[tb]
	\centering
    \includegraphics[clip, scale=0.58]{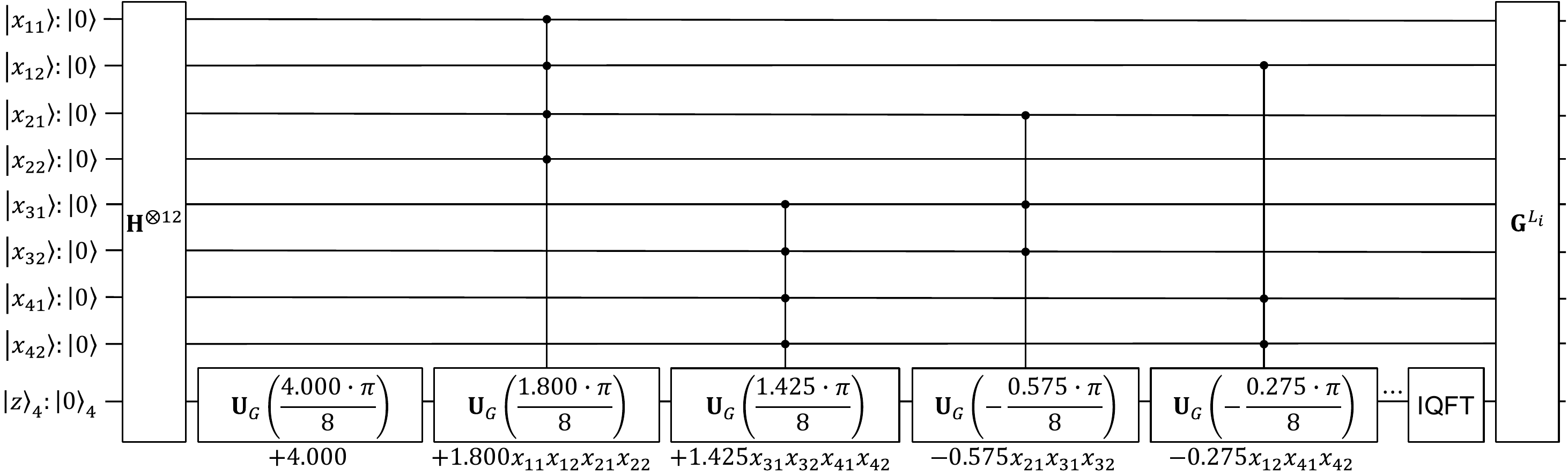}
	\caption{Quantum circuit corresponding to objective function \eqref{eq:E_prop_d}.\label{fig:circuit}}
\end{figure*}
Fig.~\ref{fig:circuit} shows a quantum circuit corresponding to \eqref{eq:E_prop_d}, where we have $n'=8$ and $m'=4$ qubits.
As shown in Fig.~\ref{fig:circuit}, the Hadamard gate $\mathbf{H}^{\otimes 12}$ at the beginning creates an equal superposition state.
The unitary operators such as $\mathbf{U}_{\mathrm{G}}(4.000 \cdot \pi / 8)$ and $\mathbf{U}_{\mathrm{G}}(5.505 \cdot \pi / 8)$ correspond to the coefficients $4.000$ and $5.505$.
Some unitary operators are controlled depending on the associated binary variables.

\section{Algebraic and Numerical Analysis\label{sec:comp}}
In this section, we analyze both conventional and proposed formulations in terms of (1) the number of binary variables, (2) the number of qubits, (3) the number of gates, and (4) query complexity.
The numbers of qubits and gates are derived in an algebraic manner and the corresponding asymptotic orders are given using the big-Omicron notation $O(\cdot)$ of \cite{knuth1976big}.

\subsection{Number of Binary Variables\label{subsec:nbin}}
The number of binary variables determines the size of the search space and the rate of convergence to the optimal solution.
In the conventional QUBO formulation \eqref{eq:E}, the number of binary variables is
\begin{align}
    n = \NAP  \NCH. \label{eq:n}
\end{align}
By contrast, in the proposed HUBO formulation \eqref{eq:E'}, the number of binary variables is
\begin{align}
    n'=\NAP \NB = \NAP\lceil \log_{2} \NCH \rceil, \label{eq:n'}
\end{align}
which is valid for both ascending and descending binary encodings.
Since the search space size of CAP is $(\NCH)^{\NAP}$, the proposed formulation becomes optimal if $\NCH$ is a power of two.

In the conventional studies, the HUBO problems have been solved by transforming the objective function to a quadratic form \cite{anthony2017quadratic}; this is termed \textit{quadratization}.
A representative quadratization approach \cite{rosenberg1975reduction} replaces a product of two binary variables with a single auxiliary binary variable and adds a constraint term to the objective function.
For example, if we have the HUBO function $f(x) = x_1 x_2 x_3$, the product $x_1 x_2$ is replaced with a new auxiliary variable $y$, and the following term,
\begin{align}
    x_1 x_2 - 2 x_1 y - 2 x_2 y + 3 y, \label{eq:exquadterm}
\end{align}
is added to the original objective function so that the constraint $x_1 x_2 = y$ holds.
Here, it is crucial to multiply \eqref{eq:exquadterm} by a scaling factor, which requires additional parameter tuning.
If the objective function is of order four or higher, the above replacement is repeated in a recursive manner.
Following the quadratization approach \cite{rosenberg1975reduction}, the proposed HUBO function can be transformed into a quadratic form that has
\begin{align}
    n''=\NAP \NB + \NAP \sum_{r=2}^{\NB} {\NB \choose r} = \NAP (2^{\NB} - 1)
    \label{eq:n''}
\end{align}
binary variables.

From \eqref{eq:n} and \eqref{eq:n'}, $n' < n$ holds for $\NCH \geq 2$, and from \eqref{eq:n'} and \eqref{eq:n''}, $n' < n''$ holds for $\NCH \geq 3$.
Hence, the proposed formulation achieves the minimum among the considered formulations.
Additionally, from \eqref{eq:n} and \eqref{eq:n''}, $n \leq n''$ holds in most cases.
However, surprisingly, if $\NCH$ is a power of two, $n > n'' = n - \NAP$ holds.

\begin{figure}[tb]
	\centering
    \includegraphics[clip, scale=0.68]{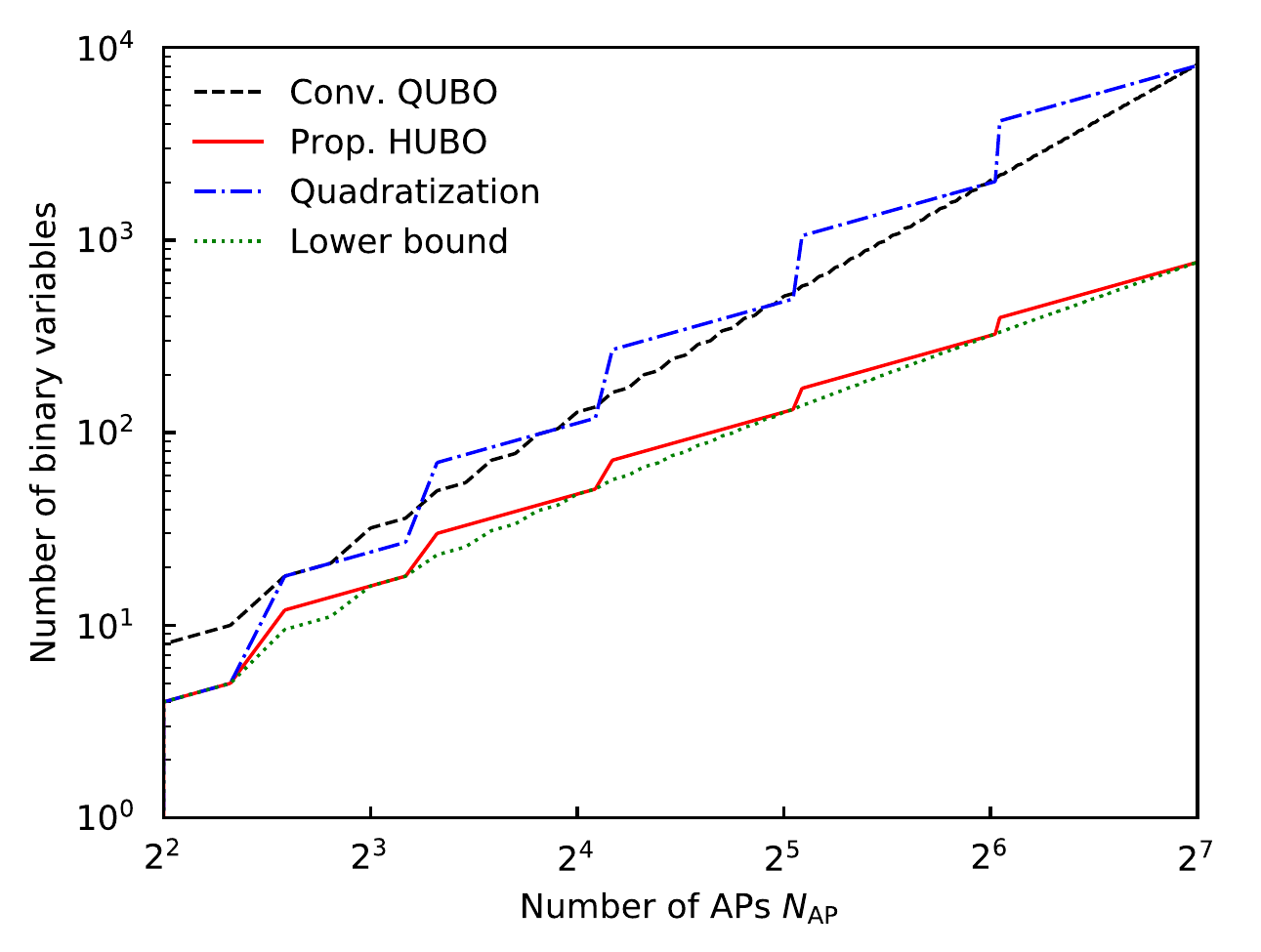}
	\caption{Actual numbers of binary variables.\label{fig:val}}
\end{figure}
In Fig.~\ref{fig:val}, we plot the actual numbers of binary variables required by the conventional QUBO, the proposed HUBO, and the QUBO transformed from HUBO (quadratization), where we fixed $\NCH = \lfloor \NAP / 2 \rfloor$ for illustration.
As shown in Fig.~\ref{fig:val}, the number of binary variables for the proposed HUBO was close to the lower bound, $\log_2 C_{\mathrm{CAP}}$, and was much smaller than for the other formulations.
Additionally, when $(\NAP, \NCH) = (128, 64)$, the QUBO formulation transformed from HUBO required $8064$ binary variables, which was less than the $8192$ of the conventional QUBO.
Here, the number of binary variables was reduced by $8192 - 8064 = 128 =\NAP$, indicating that the search space is reduced by $1/2^{128}$.

\subsection{Number of Qubits}
As introduced in Section~\ref{sec:gas}, GAS \cite{gilliam2021grover,norimoto2023quantum} requires $n + m$ qubits, where $n$ denotes the number of binary variables and $m$ denotes the number of qubits used for encoding the value of the objective function.
Since $m$ must satisfy inequalities \eqref{eq:m1} and \eqref{eq:m2}, we investigate the minimum and maximum objective functions $E(x)$ and $E'(x)$, where the ascending binary encoding is considered for the proposed HUBO.\footnote{The proposed ascending and descending encodings lead to the same minimum and maximum values.}

According to \eqref{eq:E}, in the original QUBO formulation, the minimum of the objective function is
\begin{align}
0 \leq \text{min}(E(x)) < D_{\mathrm{sum}},
\end{align}
and the maximum is
\begin{align}
\text{max}(E(x)) &=\NCH D_{\mathrm{sum}} + w \NAP(\NCH - 1)^2, \label{eq:max}
\end{align}
where we have
\begin{align}
D_{\mathrm{sum}} = \sum_{i=1}^{\NAP-1} \sum_{k=i+1}^{\NAP} D_{ik}.
\end{align}
Then, the number of qubits required to encode $E(x)$ can be calculated as
\begin{align}
m = \lceil \log_{2} (\text{max}(E(x)))\rceil +1, \label{eq:m}
\end{align}
and the total number of required qubits is
\begin{align}
&n + m \nonumber \\
= & \NAP \NCH +  \left\lceil \log_{2} \left( \NCH D_{\mathrm{sum}} + w \NAP(\NCH - 1)^2 \right) \right\rceil +1 \nonumber \\
=& O(\NAP \NCH), \label{eq:nm}
\end{align}
which indicates that the number of binary variables has the dominant effect.

In the proposed HUBO formulation, from \eqref{eq:E_prop_a} and \eqref{eq:E_prop_d}, the minimum of the objective function is
\begin{align}
0 \leq \text{min}(E'_{\mathrm{a}}(x)) = \text{min}(E'_{\mathrm{d}}(x)) < D_{\mathrm{sum}},
\end{align}
and the maximum is
\begin{align}
\text{max}(E'_{\mathrm{a}}(x)) = \text{max}(E'_{\mathrm{d}}(x))
= \sum_{i=1}^{\NAP-1} \sum_{k=i+1}^{\NAP} D_{ik} = D_{\mathrm{sum}}. \label{eq:max'}
\end{align}
Then, the number of qubits required to encode $E'_\mathrm{a}(x)$ or $E'_\mathrm{d}(x)$ can be calculated as
\begin{align}
m' = \lceil \log_{2} (\text{max}(E'_{\mathrm{a}}(x)))\rceil + 1, \label{eq:m'}
\end{align}
and the total number of required qubits is
\begin{align}
n' + m' &=  \NAP \lceil \log_{2} \NCH \rceil + \left\lceil \log_{2} D_{\mathrm{sum}} \right \rceil +1 \nonumber \\
&= O(\NAP \lceil \log \NCH \rceil). \label{eq:nm'}
\end{align}
Since $\NCH > \lceil \log_2{\NCH} \rceil$ and $w > 0$ hold, the number of required qubits for the proposed HUBO is less than that for the conventional QUBO, i.e., $n' < n$ and $m' < m$.

To simplify the comparison, we fix both the distance $D_{ik}$ and the penalty factor $w$ to $1$, and fix the number of channels to $\NCH = \lfloor \NAP / 2 \rfloor$.
In this simplified case, $D_{\mathrm{sum}}$ becomes $\NAP (\NAP - 1) / 2$,
Then, the conventional formulation requires $n + m = O(\NAP^2)$ qubits,
while the proposed formulation requires $n' + m' = O(\NAP \lceil \log \NAP \rceil)$ qubits.
\begin{figure}[tb]
	\centering
    \includegraphics[clip, scale=0.68]{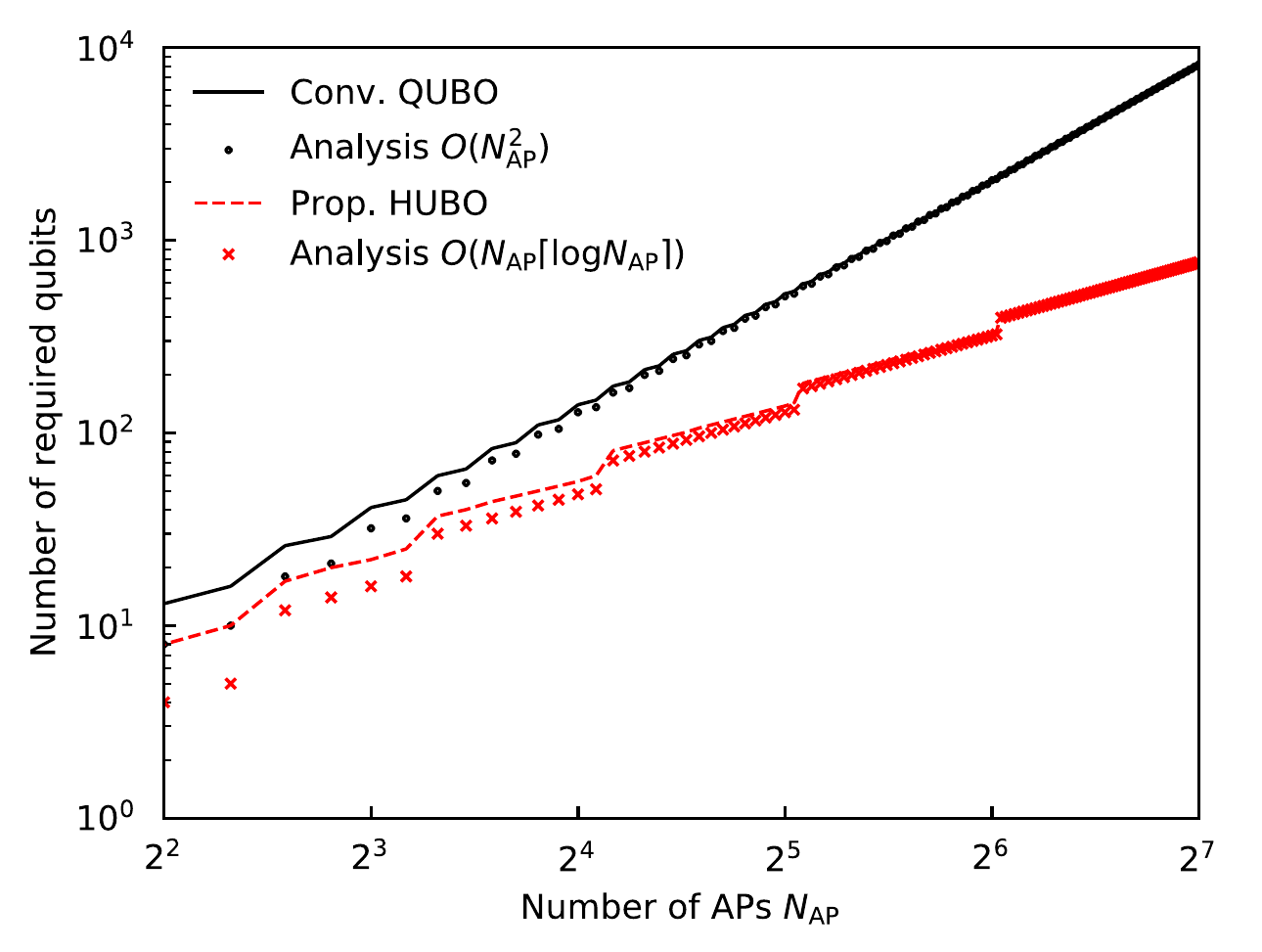}
	\caption{Numbers of required qubits, with lines indicating actual values and markers indicating theoretical values.\label{fig:qubit}}
\end{figure}
In Fig.~\ref{fig:qubit}, we plot the derived order as well as the actual number of qubits required by constructed circuits.
As shown in Fig.~\ref{fig:qubit}, both analytical and numerical values basically corresponded, demonstrating the accuracy of our analysis.
The proposed formulation succeeded in reducing the number of required qubits without exception.
Although the effect of the non-negligible terms is large when $\NAP$ is small, the quadratic speedup by GAS is particularly dominant when the problem size is large.

\subsection{Number of Gates}
The number of quantum gates is an important evaluation metric because it determines the feasibility of quantum circuits and quantum algorithms.
As in the conventional studies \cite{gilliam2021grover,norimoto2023quantum},
we construct a specific quantum circuit using GAS and count the number of quantum gates.
In the quantum circuit of GAS, the most complex part is the gates corresponding to $\mathbf{A}_{y_i}$.
In the following, we analyze the number of gates used in $\mathbf{A}_{y_i}$.

In the conventional QUBO formulation, from \eqref{eq:nm}, the number of H gates is
\begin{align}
    G_{\mathrm{H}} = n+m = O(\NAP \NCH).
\end{align}
The numbers of R and controlled R (CR) gates can be calculated using \eqref{eq:m}.
The number of terms of the first order corresponds to the number of 1-CR gates, which is calculated as
\begin{align}
    G_{\textrm{1-CR}} =& \NAP \NCH \beta \nonumber \\
    =& O(\NAP \NCH \log (\NAP^2 \NCH + \NAP\NCH^2)), \label{eq:QUBO1CR}
\end{align}
where we have
\begin{align}
\beta = \left\lceil \log_{2} \left(\NCH {\NAP \choose 2} + \NAP(\NCH - 1)^2 \right)  \right\rceil +1 \nonumber \\
= \log_{2} \left(\NCH {\NAP \choose 2} + \NAP(\NCH - 1)^2 \right) + O(1).
\end{align}
Similarly, we count the number of quadratic terms in the objective function. The corresponding number of 2-CR gates is calculated as
\begin{align}
&G_{\textrm{2-CR}} = \left(\NCH \binom{\NAP}{2} + \NAP \binom{\NCH}{2} \right) \cdot \beta \nonumber \\
&=\frac{\NAP\NCH(\NAP+\NCH-2)}{2} \cdot \beta \nonumber \\
&= O(\NAP\NCH(\NAP+\NCH) \log (\NAP^2 \NCH + \NAP\NCH^2)).
\label{eq:QUBO2CR}
\end{align}

In the proposed HUBO formulation, from \eqref{eq:nm'}, the number of H gates is
\begin{align}
    G_{\textrm{H}}' = n' + m' = O(\NAP \log \NCH).
\end{align}
The numbers of R and CR gates are calculated using \eqref{eq:m'}.
Specifically, the number of 1-CR gates is calculated as
\begin{align}
    G_{\textrm{1-CR}}' &= \NAP \NB \cdot \beta' = \NAP \left\lceil \log_{2} \NCH \right \rceil \cdot \beta' \nonumber \\
    &=O(\NAP \log \NAP \log \NCH) \label{eq:HUBO1CR}
\end{align}
where we have
\begin{align}
    \beta' = \left\lceil \log_{2} {\NAP \choose 2} \right\rceil +1 = O(\log \NAP).
\end{align}
Similarly, the number of 2-CR gates is
\begin{align}
G_{\textrm{2-CR}}' &= \binom{\NAP \NB}{2} \cdot \beta'
= \frac{\NAP \NB (\NAP \NB - 1)}{2} \cdot \beta' \nonumber \\
&= O(\NAP^2 \log \NAP (\log \NCH) ^2). \label{eq:HUBO2CR}
\end{align}
Additionally, the number of $k$-CR gates is
\begin{align}
G_{k\textrm{-CR}}' &= \left\{ \binom{\NAP}{2} \binom{2\NB}{k} - \NAP(\NAP-2) \binom{\NB}{k} \right\} \cdot \beta' \nonumber \\
&= O\left(\NAP^2 \log \NAP \frac{2^k (\log \NCH)^k}{k!} \right) \label{eq:k1}
\end{align}
for $3 \leq k \leq \NB$, while it is
\begin{align}
G_{k\textrm{-CR}}' &= \binom{\NAP}{2} \binom{2\NB}{k} \cdot \beta' \nonumber \\
&= O\left(\NAP^2 \log \NAP \frac{2^k (\log \NCH)^k}{k!} \right) \label{eq:k2}
\end{align}
for $\NB < k \leq 2\NB$.

\begin{table}[tbp]
	\centering
	\caption{Number of quantum gates required by $\mathbf{A}_{y_i}$. \label{table:gate}}
	\begin{tabular}{lll}
	    \hline
	    Gate&Conventional QUBO&Proposed HUBO\\
	    \hline
	    H&$O(\NAP^2)$&$O(\NAP \log \NAP)$\\
	    R&$O(\log \NAP)$&$O(\log \NAP)$\\
	    1-CR&$O(\NAP^2 \log \NAP)$&$O(\NAP \left(\log \NAP \right)^2)$\\
	    2-CR&$O(\NAP^3 \log \NAP)$&$O(\NAP^2 (\log \NAP)^3)$\\
	    $k$-CR&$0$&$O\left(\NAP^2 (\log \NAP)^{k+1} / k! \right)$\\
	    IQFT&$1$&$1$\\
	    \hline
	\end{tabular}
\end{table}
From the results of the above analysis, the number of quantum gates required by the state preparation operator $\mathbf{A}_{y_i}$ is summarized in Table~\ref{table:gate}, where the number of channels is set to $\NCH = \lfloor \NAP / 2 \rfloor$ for simplicity.
Compared with the conventional formulation, the proposed formulation reduces the number of H gates, while it increases the total number of 1-CR, 2-CR, $\cdots$, $(2 \NB)$-CR gates.
The increase in the number of required gates is a major drawback of the proposed formulation.

\begin{figure}[tb]
	\centering
    \includegraphics[clip, scale=0.655]{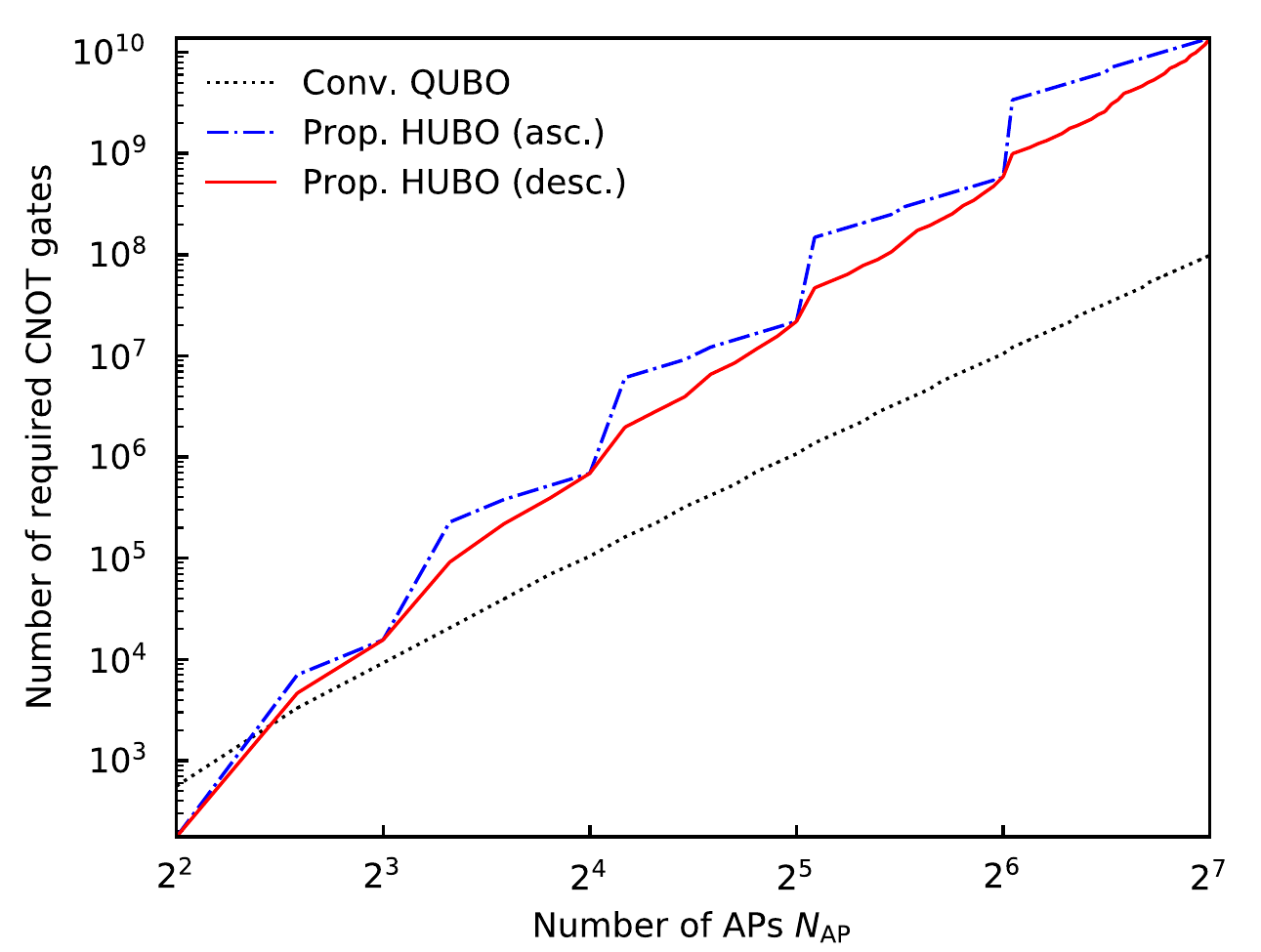}
	\caption{Actual numbers of required CNOT gates.\label{fig:g-comp}}
\end{figure}
To overcome this drawback, in Section~\ref{sec:prop}, we proposed the descending binary encoding, which assigns a binary number to the channel index in descending order.
In Fig.~\ref{fig:g-comp}, we plot the effects of ascending and descending encodings, where the number of APs was increased from $\NAP = 2^2$ to $2^7$ and the number of channels was $\NCH = \lfloor \NAP / 2 \rfloor$.
%
%
In actual quantum computers, the quantum circuit is decomposed into elementary gates \cite{nielsen2010quantum}. After decomposition, the number of controlled NOT (CNOT) or T gates has been used to evaluate the complexity of quantum circuit \cite{gidney2021how,tezuka2022grover}, termed as gate complexity.
A 1-CR gate can be decomposed into elementary gates including two CNOT gates \cite{nielsen2010quantum}.
Using Maslov's approach \cite{maslov2016advantages} and relative-phase Toffoli gates, a decomposed $n$-CR gate has $6(n-1)$ CNOT gates. Then, we counted the total number of CNOT gates for both encodings.\footnote{Note that this decomposition requires ancilla qubits of $\mathrm{deg}(E(x))-1$ for QUBO and $\mathrm{deg}(E'(x))-1$ for HUBO, where $\mathrm{deg}(\cdot)$ is the degree of the objective function. This increase in ancilla qubits is negligible with respect to the number of required qubits $n+m$.}
As shown in Fig.~\ref{fig:g-comp}, the number of CNOT gates was reduced by the proposed descending encoding.
Specifically, the difference between ascending and descending encodings was particularly large when $\log_2 \NAP \neq \lceil \log_2 \NAP \rceil$.
We observed the same trend in our comparison of the number of T gates. Note that a particular gate decomposition method does not influence the relative advantages of different formulations.
As the problem size increases, the gate complexity also increases correspondingly, which can lead to a fatal issue when using an actual quantum computer with gates that have finite fidelity.

\subsection{Query Complexity}
Finally, we investigate the query complexities of GAS that are required by solving the conventional QUBO and the proposed HUBO formulations.
As clarified in Section~\ref{subsec:nbin}, the conventional QUBO formulation requires $n=\NAP \NCH$ binary variables, while the proposed HUBO formulation requires $n'= \NAP\lceil \log_{2} \NCH \rceil$ binary variables.
Since GAS provides a quadratic speedup, the two query complexities are $\sqrt{2^n} = 2^{\NAP \NCH/2}$ and
\begin{align}
    (\NCH)^{\frac{\NAP}{2}}
    \leq
    \sqrt{2^{n'}}
    <
    (\sqrt{2}\NCH)^{\frac{\NAP}{2}},
\end{align}
respectively.
Obviously, the reduction rate is maximized if $\NAP = \NCH$, and it becomes smaller as $\NCH$ decreases.
\begin{figure}[tb]
	\centering
    \includegraphics[clip, scale=0.68]{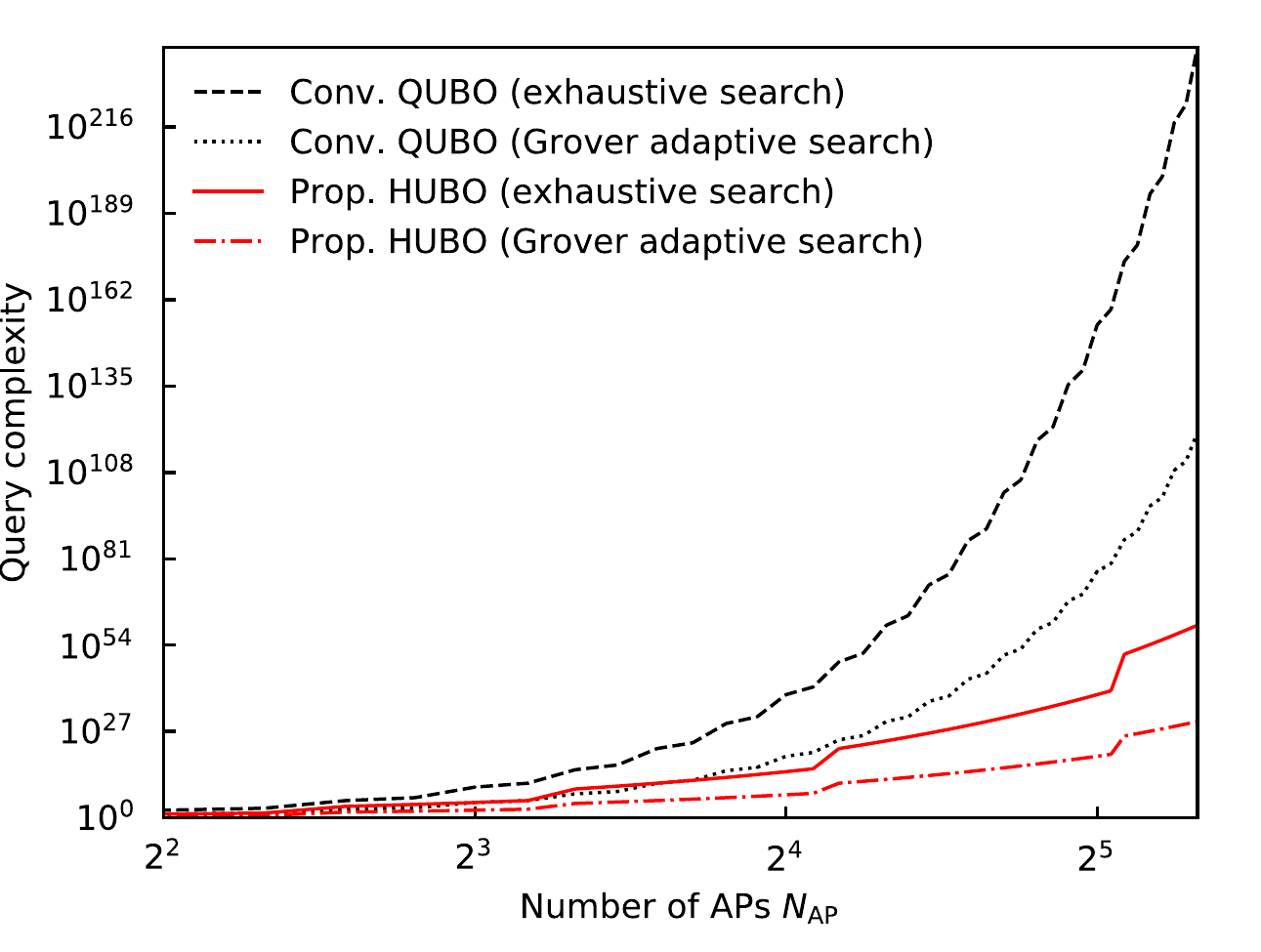}
	\caption{Expected query complexity in the quantum domain.\label{fig:query}}
\end{figure}
In Fig.~\ref{fig:query}, we show expected query complexities of both formulations, where the number of channels was $\NCH = \lfloor \NAP / 2 \rfloor$.
As a reference, the query complexity for the exhaustive search is also shown, which is equivalent to the average number of objective function evaluations.
As shown in Fig.~\ref{fig:query}, the classic exhaustive search exhibited impractical complexities for both conventional and proposed formulations, and is infeasible on classical computers.
By contrast, it can be expected that the proposed HUBO formulation using GAS significantly reduces the query complexity, and the reduction rate improves as the problem size increases.

\begin{figure}[tb]
	\centering
	\subfigure[Query complexity in the classical domain.]{
		\includegraphics[clip, scale=0.68]{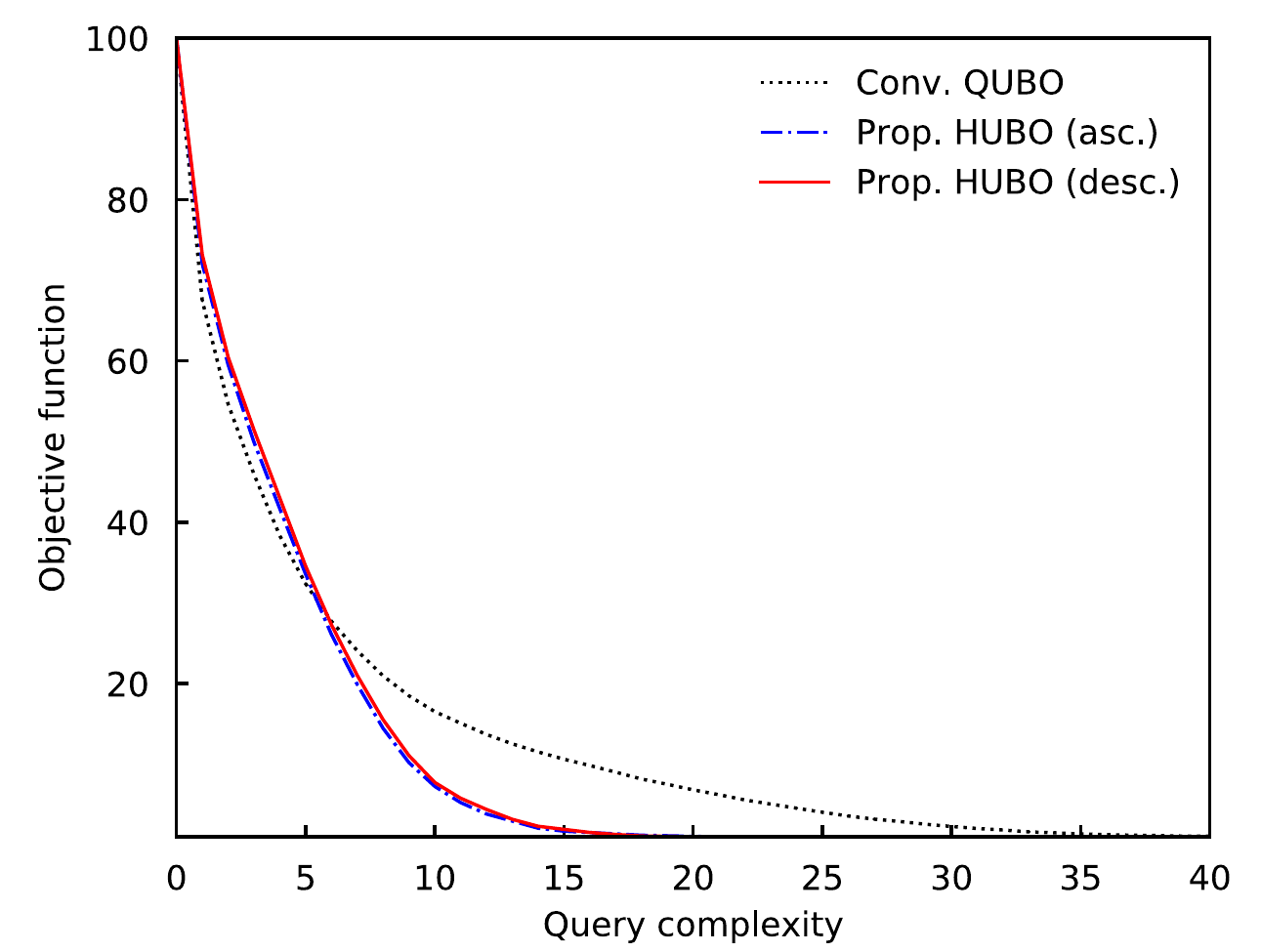}
	}
	\subfigure[Query complexity in the quantum domain.]{
		\includegraphics[clip, scale=0.68]{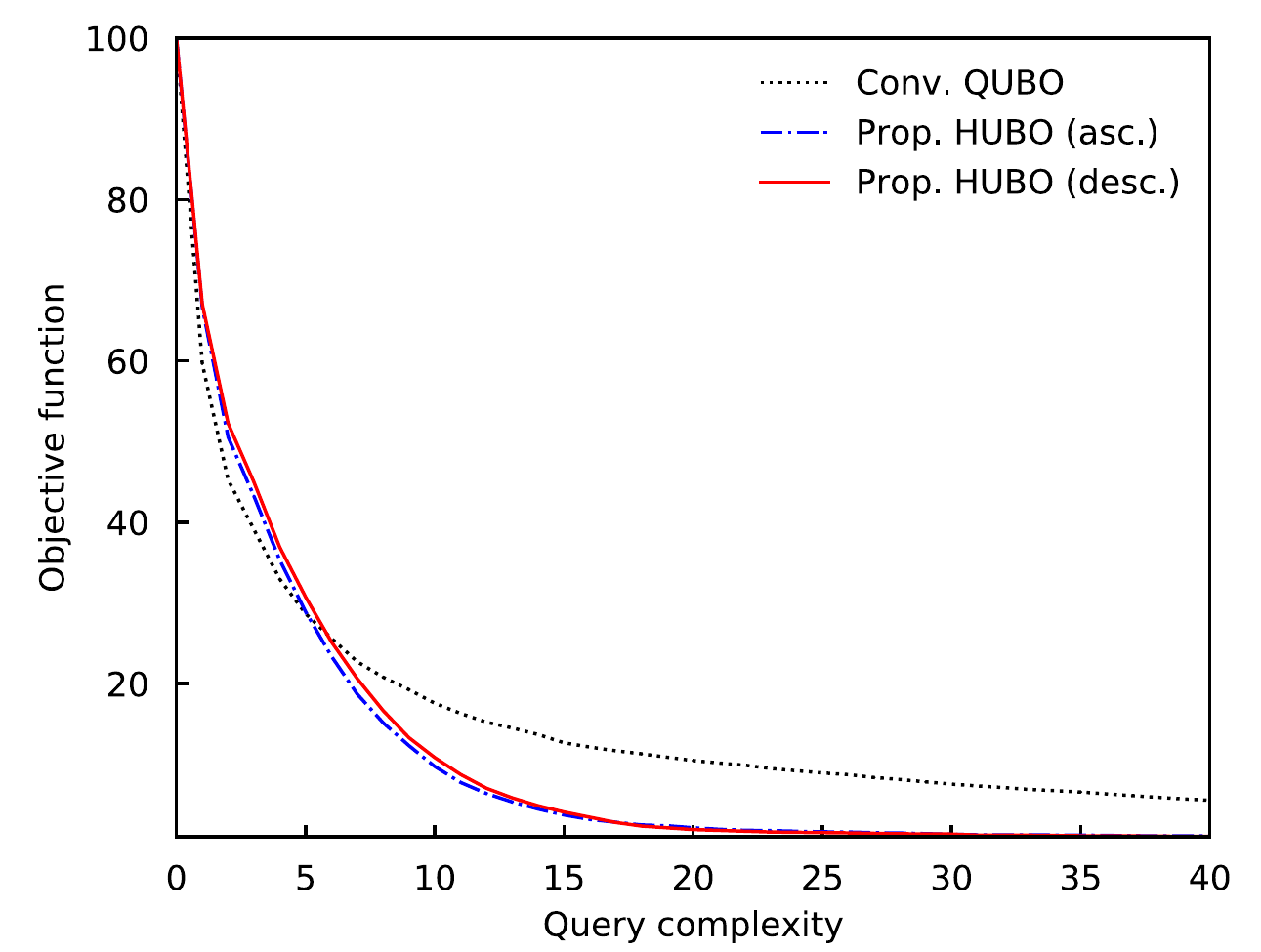}
	}
	\caption{Query complexities required by ideal GAS.\label{fig:ite}}
\end{figure}
In Fig.~\ref{fig:ite}, we show the query complexities for the conventional and proposed formulations, which are given in Appendices~\ref{app:exqubo} and \ref{app:exhubo}, respectively.
Here, we set $\NAP = 4$ and $\NCH = 3$, evaluated the query complexity in the classical and quantum domains, and normalized the objective function values from $0$ to $100$.
Our intention is to investigate how each formulation affects the query complexity, and due to the time-consuming nature of quantum simulations, we considered here an ideal GAS using the integer approximation \cite{gilliam2021grover} where the number of qubits to encode an objective function value, $m$, is sufficiently large.
In all cases, the proposed HUBO formulation reached the optimal solution with reduced query complexity compared with the conventional QUBO formulation, which is the main advantage of the proposed formulation.
This gap is expected to become large as the problem size increases, as given in Fig.~\ref{fig:query}.
Additionally, as expected, both ascending and descending encodings exhibited the same query complexities.
Each objective function includes real-valued coefficients. That is, when $m$ is limited to a practical value, a degradation in query complexity is inferred from the conventional study \cite{norimoto2023quantum} dealing with the real-valued GAS. However, this negative effect applies equally to the conventional and proposed formulations, and it does not affect their relative comparisons.

\section{Conclusions \label{sec:conc}}
In this paper, we formulated the NP-hard wireless CAP as a HUBO problem.
To solve CAP, the conventional QUBO formulation uses the one-hot encoding of channel indices, while the proposed HUBO uses the ascending or descending binary encoding of the indices. The proposed formulation was analyzed in terms of the number of binary variables, the number of qubits, the number of gates, and the query complexity. When using GAS, we found that the proposed HUBO significantly reduced the number of qubits and the query complexity compared with the conventional QUBO formulation. This advantage was achieved at the cost of increased number of gates; however, the proposed descending binary encoding also succeeded in reducing the number of gates.

\section*{Acknowledgement}
The authors would like to thank Prof. Naoki Yamamoto, Keio University, Japan for providing valuable comments.

\appendices
\section{Example of QUBO Formulation\label{app:exqubo}}
We describe the QUBO formulation for the case of $\NAP = 4$ and $\NCH = 3$, which was used in Fig.~\ref{fig:ite}.
We assume that the physical distances $d_{iu}$ between the $i$th AP and $u$th UT for $1 \leq i \leq \NAP$ and $1 \leq u \leq 2\NAP$ in Fig.~\ref{fig:model} are
\begin{align}
&\begin{bmatrix}
d_{11} & d_{12} & d_{13} & d_{14} & d_{15} & d_{16} & d_{17} & d_{18} \\
d_{21} & d_{22} & d_{23} & d_{24} & d_{25} & d_{26} & d_{27} & d_{28} \\
d_{31} & d_{32} & d_{33} & d_{34} & d_{35} & d_{36} & d_{37} & d_{38} \\
d_{41} & d_{42} & d_{43} & d_{44} & d_{45} & d_{46} & d_{47} & d_{48}
\end{bmatrix} \nonumber \\
=&
\begin{bmatrix}
1 & 1 & 2 & 4 & 3 & 5 & 5 & 8\\
5 & 4 & 1 & 1 & 5 & 4 & 2 & 4\\
6 & 5 & 2 & 5 & 1 & 1 & 4 & 6\\
10 & 8 & 5 & 2 & 5 & 3 & 1 & 1
\end{bmatrix}.
\label{eq:div}
\end{align}
Then, the scalar constant $C_{ik}$ \eqref{eq:Cik} for $1 \leq i < k \leq \NAP$ in Fig.~\ref{fig:model} is calculated as
{
\footnotesize
\begin{align}
C_{12} &= -\log_2\left(1 + \frac{1^{-1}+1^{-1}}{2^{-1}+4^{-1}}\right) - \log_2\left(1 + \frac{1^{-1}+1^{-1}}{ 5^{-1}+4^{-1}} \right) = -4.319 \nonumber \\
C_{13} &= -\log_2\left(1 + \frac{1^{-1}+1^{-1}}{3^{-1}+5^{-1}}\right) - \log_2\left(1 + \frac{1^{-1}+1^{-1}}{ 6^{-1}+5^{-1}} \right) = -4.938 \nonumber \\
C_{14} &= -\log_2\left(1 + \frac{1^{-1}+1^{-1}}{5^{-1}+8^{-1}}\right) - \log_2\left(1 + \frac{1^{-1}+1^{-1}}{10^{-1}+8^{-1}} \right) = -6.145 \nonumber \\
C_{23} &= -\log_2\left(1 + \frac{1^{-1}+1^{-1}}{5^{-1}+4^{-1}}\right) - \log_2\left(1 + \frac{1^{-1}+1^{-1}}{ 2^{-1}+5^{-1}} \right) = -4.392 \nonumber \\
C_{24} &= -\log_2\left(1 + \frac{1^{-1}+1^{-1}}{2^{-1}+4^{-1}}\right) - \log_2\left(1 + \frac{1^{-1}+1^{-1}}{ 5^{-1}+2^{-1}} \right) = -3.822 \nonumber \\
C_{34} &= -\log_2\left(1 + \frac{1^{-1}+1^{-1}}{4^{-1}+6^{-1}}\right) - \log_2\left(1 + \frac{1^{-1}+1^{-1}}{ 5^{-1}+3^{-1}} \right) = -4.784
\label{eq:cik_ex}
%
%
\end{align} \normalsize }where we have the attenuation coefficient $\alpha = 1$ for simplicity.
Now, we have $C_{\mathrm{min}} = -6.145$.
The constants $D_{ik} = C_{ik} - C_{\mathrm{min}} + \epsilon$ \eqref{eq:Dik} with $\epsilon=0.01$ are calculated as $D_{12} = 1.835, D_{13} = 1.216, D_{14} = 0.010, D_{23} = 1.762, D_{24} = 2.333,$ and $D_{34} = 1.371$.
In this case, from \eqref{eq:E}, the conventional objective function is
\begin{align}
E(x) =+&1.835(x_{11}x_{21} + x_{12}x_{22} + x_{13}x_{23}) \nonumber \\
      +&1.216(x_{11}x_{31} + x_{12}x_{32} + x_{13}x_{33}) \nonumber \\
      +&0.010(x_{11}x_{41} + x_{12}x_{42} + x_{13}x_{43}) \nonumber \\
      +&1.762(x_{21}x_{31} + x_{22}x_{32} + x_{23}x_{33}) \nonumber \\
      +&2.333(x_{21}x_{41} + x_{22}x_{42} + x_{23}x_{43}) \nonumber \\
      +&1.371(x_{31}x_{41} + x_{32}x_{42} + x_{33}x_{43}) \nonumber \\
      +&(x_{11} + x_{12} + x_{13} - 1)^2 + (x_{21} + x_{22} + x_{23} - 1)^2 \nonumber \\
      +&(x_{31} + x_{32} + x_{33} - 1)^2 + (x_{41} + x_{42} + x_{43} - 1)^2.
      \label{eq:E_conv}
\end{align}
The optimal solution is a set of binary vectors
\begin{align}
[x_{11} ~ x_{12} ~ x_{13}] &= [0 ~ 1 ~ 0] \rightarrow 2, \nonumber \\
[x_{21} ~ x_{22} ~ x_{23}] &= [1 ~ 0 ~ 0] \rightarrow 1, \nonumber \\
[x_{31} ~ x_{32} ~ x_{33}] &= [0 ~ 0 ~ 1] \rightarrow 3, ~ \mathrm{and} \nonumber \\
[x_{41} ~ x_{42} ~ x_{43}] &= [0 ~ 1 ~ 0] \rightarrow 2, \label{eq:solu_qubo}
\end{align}
which are one-hot vectors.
Solution \eqref{eq:solu_qubo} indicates that the first, second, third, and fourth APs use the second, first, third, and second channels, respectively.

\section{Example of HUBO Formulations \label{app:exhubo}}
\begin{figure*}[t!]
\begin{align}
      E_{\mathrm{a}}'(x) =
       +&1.835\{(1-x_{11})(1-x_{12})(1-x_{21})(1-x_{22}) + (1-x_{11})x_{12}(1-x_{21})x_{22} + x_{11}(1-x_{12})x_{21}(1-x_{22})\} \nonumber \\
       +&1.216\{(1-x_{11})(1-x_{12})(1-x_{31})(1-x_{32})+ (1-x_{11})x_{12}(1-x_{31})x_{32} + x_{11}(1-x_{12})x_{31}(1-x_{32})\} \nonumber \\
       +&0.010\{(1-x_{11})(1-x_{12})(1-x_{41})(1-x_{42})+ (1-x_{11})x_{12}(1-x_{41})x_{42} + x_{11}(1-x_{12})x_{41}(1-x_{42})\} \nonumber \\
       +&1.762\{(1-x_{21})(1-x_{22})(1-x_{31})(1-x_{32})+ (1-x_{21})x_{22}(1-x_{31})x_{32} + x_{21}(1-x_{22})x_{31}(1-x_{32})\} \nonumber \\
       +&2.333\{(1-x_{21})(1-x_{22})(1-x_{41})(1-x_{42})+ (1-x_{21})x_{22}(1-x_{41})x_{42} + x_{21}(1-x_{22})x_{41}(1-x_{42})\} \nonumber \\
       +&1.371\{(1-x_{31})(1-x_{32})(1-x_{41})(1-x_{42})+ (1-x_{31})x_{32}(1-x_{41})x_{42} + x_{31}(1-x_{32})x_{41}(1-x_{42})\} \nonumber \\
      + &x_{11}x_{12} + x_{21}x_{22} + x_{31}x_{32} + x_{41}x_{42} \label{eq:E_prop_a} \\
      E_{\mathrm{d}}'(x) =
       +&1.835\{x_{11}x_{12} x_{21}x_{22} + x_{11}(1-x_{12})x_{21}(1-x_{22})+ (1-x_{11})x_{12}(1-x_{21})x_{22}\} \nonumber \\
       +&1.216\{x_{11}x_{12} x_{31}x_{32} + x_{11}(1-x_{12})x_{31}(1-x_{32})+ (1-x_{11})x_{12}(1-x_{31})x_{32}\} \nonumber \\
       +&0.010\{x_{11}x_{12} x_{41}x_{42} + x_{11}(1-x_{12})x_{41}(1-x_{42})+ (1-x_{11})x_{12}(1-x_{41})x_{42}\} \nonumber \\
       +&1.762\{x_{21}x_{22} x_{31}x_{32} + x_{21}(1-x_{22})x_{31}(1-x_{32})+ (1-x_{21})x_{22}(1-x_{31})x_{32}\} \nonumber \\
       +&2.333\{x_{21}x_{22} x_{41}x_{42} + x_{21}(1-x_{22})x_{41}(1-x_{42})+ (1-x_{21})x_{22}(1-x_{41})x_{42}\} \nonumber \\
       +&1.371\{x_{31}x_{32} x_{41}x_{42} + x_{31}(1-x_{32})x_{41}(1-x_{42})+ (1-x_{31})x_{32}(1-x_{41})x_{42}\} \nonumber \\
       -&(1-x_{11})(1-x_{12}) + (1-x_{21})(1-x_{22})
      + (1-x_{31})(1-x_{32}) + (1-x_{41})(1-x_{42}) \label{eq:E_prop_d}
\end{align}
\hrulefill
\vspace*{4pt}
\end{figure*}
We describe the HUBO formulations for the case of $\NAP = 4$ and $\NCH = 3$ used in Fig.~\ref{fig:ite}.
Specifically, the proposed objective function using ascending encoding is given in \eqref{eq:E_prop_a}, while that using descending encoding is given in \eqref{eq:E_prop_d}.
As given, the number of terms is reduced, i.e., from 67 to 55.
The optimal solution for ascending encoding \eqref{eq:E_prop_a} is a set of binary vectors
\begin{align}
[x_{11} ~ x_{12}] &= [0 ~ 1] \rightarrow 2, \nonumber \\
[x_{21} ~ x_{22}] &= [0 ~ 0] \rightarrow 1, \nonumber \\
[x_{31} ~ x_{32}] &= [1 ~ 0] \rightarrow 3, ~ \mathrm{and} \nonumber \\
[x_{41} ~ x_{42}] &= [0 ~ 1] \rightarrow 2, \label{eq:solu_hubo_a}
\end{align}
while the optimal solution for descending encoding \eqref{eq:E_prop_d} is a set of binary vectors
\begin{align}
[x_{11} ~ x_{12}] &= [1 ~ 0] \rightarrow 2, \nonumber \\
[x_{21} ~ x_{22}] &= [1 ~ 1] \rightarrow 1, \nonumber \\
[x_{31} ~ x_{32}] &= [0 ~ 1] \rightarrow 3, ~ \mathrm{and} \nonumber \\
[x_{41} ~ x_{42}] &= [1 ~ 0] \rightarrow 2. \label{eq:solu_hubo_d}
\end{align}
Both solutions indicate the same situation as that given in \eqref{eq:solu_qubo}.

\footnotesize{
	\bibliographystyle{IEEEtran}
	\bibliography{main}
}

\begin{IEEEbiographynophoto}{Yuki~Sano}
received the B.E. degree in engineering science from Iwate University, Iwate, Japan, in 2021, and the M.E. degree in engineering science from Yokohama National University, Kanagawa, Japan, in 2023. He is currently working at Nomura Research Institute based in Tokyo.
\end{IEEEbiographynophoto}

\begin{IEEEbiographynophoto}{Masaya~Norimoto}
(Graduate Student Member, IEEE) received the B.E. degree in engineering science from Yokohama National University, Kanagawa, Japan, in 2022. He is currently pursuing the M.E. degree with the Graduate School of Engineering Science, Yokohama National University, Kanagawa, Japan. His research interests include quantum algorithms and wireless communications.
\end{IEEEbiographynophoto}

\begin{IEEEbiographynophoto}{Naoki~Ishikawa}
(Senior Member, IEEE) is an Associate Professor with the Faculty of Engineering, Yokohama National University, Kanagawa, Japan. He received the B.E., M.E., and Ph.D. degrees in electronic and information engineering from the Tokyo University of Agriculture and Technology, Tokyo, Japan, in 2014, 2015, and 2017, respectively. In 2015, he was an academic visitor with the School of Electronics and Computer Science, University of Southampton, UK. From 2016 to 2017, he was a research fellow of the Japan Society for the Promotion of Science. From 2017 to 2020, he was an assistant professor in the Graduate School of Information Sciences, Hiroshima City University, Japan. He was certified as an Exemplary Reviewer of \textsc{IEEE Transactions on Communications} in 2017 and 2021. His research interests include massive MIMO, physical layer security, and quantum speedup for wireless communications.
\end{IEEEbiographynophoto}

\end{document}